# Disorder-immune momentum band winding topology

**Authors:** Andrea Steinfurth[1]†, Sebastian Weidemann[1]†, Julia Görsch[1], Tom Sheppard[2], Hannah M. Price[2], Alexander Szameit[1]*, Joshua Feis[1]

**Affiliations:**

[1]Institute of Physics, University of Rostock; Rostock, Germany

[2]School of Physics and Astronomy, University of Birmingham; Birmingham, United Kingdom

*Corresponding author. Email: alexander.szameit@uni-rostock.de

†These authors contributed equally to this work.

**Time is the odd dimension out:** Unlike space, it follows the arrow of time, forbidding back-reflections and requiring momentum yet not energy conservation. Tailored temporal variations manipulate momentum bands and engineer waves in time. We show that momentum bands exhibit unique topology, hidden when conventionally considering energy bands: Complex momentum bands may wind, mandating topological localization at time interfaces. We observe this effect in photonic quantum walks and study it under disorder. Remarkably, unlike any known topological phenomenon, the topology is immune against arbitrarily strong disorder. Only exotic conditions through extreme spatiotemporally random non-Hermiticity can destroy it. Our findings uncover a disorder-immune type of topological physics, inviting explorations of complex momentum or energy-momentum topology with potential applications like ultrarobust lasing, temporal pulse shaping or amplification.

Conventionally, crystalline matter is defined to be periodic in space and described as well as engineered in terms of energy band structure (*1*). However, recently, rapid progress on time crystals (*2*, *3*) and time-varying media (*4*, *5*), alternatively called temporal crystals (*6*, *7*) or photonic time crystals (*8*, *9*), has broken this paradigm by engineering periodicity not in space but in time. However, causality makes time unidirectional and thus often introduces a unique asymmetry between physics on the time axis and on the spatial axes. Due to the arrow of time, back-reflections at time interfaces are forbidden and energy is not conserved (*10*). Correspondingly, the physics of time-varying matter most naturally expresses itself not in terms of energy and energy bands: Instead, time reflections conserve momentum and time-periodicity leads to a momentum band structure (*11*, *12*). The properties and engineering of momentum bands and momentum band gaps give rise to a rapidly widening range of novel and intriguing physical phenomena in time, such as time edge states (*7*, *13*, *14*), superluminal solitons (*15*) or threshold-free lasing (*16*), to name only a few. While most of these properties and effects may be inferred from or are at least hinted at by the energy band structure, certain effects remain hidden from this conventional description: While studies so far have focused on systems with real momentum bands, in those with complex momentum bands, there generally is no direct connection between the momentum and energy band structure. Correspondingly, considerations of the complex momentum may unearth previously hidden features. Notably, from the viewpoint of topological physics, as complex functions momentum bands may potentially exhibit winding and correspondingly non-trivial topology.

In the following work, we explore the properties of complex momentum bands, finding that they may indeed exhibit winding and defining their associated topological invariant. We further construct models exhibiting topologically non-trivial momentum band winding topology based on quantum walks. Based on this, we present a photonic experimental implementation of such a quantum walk model. We demonstrate that the momentum band winding topology leads to temporal localization at a time interface akin to the non-Hermitian skin effect but in time, as illustrated in Fig. 1. Finally, we examine the behavior of this effect under disorder in both simulation as well as experiment and find that it is topologically protected. Remarkably and uniquely, we observe that topology and localization are entirely immune against disorder through random potentials with arbitrary strength and only break down under extreme random non-Hermiticity.

*Theory.* The spectrum of energy eigenvalues, together with the set of Floquet-Bloch energy eigenstates, has conventionally been used to lay bare all topological properties of physical systems (*17*, *18*): Initially, in Hermitian systems, by studying the eigenstates surrounding a gap in energy eigenvalues, topological states living in that gap and localized at spatial interfaces may emerge (*19*, *20*). Later, in non-Hermitian systems, the complex energy eigenvalues could be seen to wind around a reference point: the energy point gap leading to the localization of all eigenstates at spatial interfaces, a phenomenon known as the non-Hermitian skin effect (*21*, *22*). Most recently, the notion of momentum bands and gaps, in this context typically defined as intervals of purely real and imaginary eigenvalues, respectively, was demonstrated to lead to topological states localized at interfaces in time (7, 14, 23). All these features have in common that they are readily evident from the conventional energy band structure. However, as we show here, there is topology and associated topological physics which remains hidden if only energy bands are considered: Instead of the Hamiltonian $H(k)$, which has a spectrum of complex energies for the input set of Bloch momenta $k$, we now consider the momentum operator $p(E)$ of the system, which is a function of the quasienergies $E$. As is well known (24), whereas the Hamiltonian generates the time evolution of a system, the momentum operator generates the spatial evolution. The spectrum of the momentum operator is the momentum band structure. Note that for a notion of bands to emerge the system must necessarily be periodic in time and thus energy, i.e., temporally crystalline. While for purely real energy eigenvalues momentum and energy bands coincide (7), leading to some features being observable in both, in systems with complex energies this is generally not the case. Crucially, for instance, momentum bands may exhibit a winding that is not at all reflected in the energy band structure, which remains topologically trivial. More explicitly, this winding is given by a winding number (25) defined as

$$w(k) \equiv \sum_{n=1}^{N} \int_{-\pi}^{\pi} \frac{dE}{2\pi} \partial_E \arg\left[e^{-ik_n(E)} - e^{-ik}\right], \qquad (1)$$

where $N$ is the number of momentum bands.

To construct a concrete example of a system with such momentum band winding, we consider a photonic quantum walk, governed by recursive evolution equations

$$\begin{aligned} u_x^{t+1} &= [\cos(\beta)\, u_{x+1}^t + i\sin(\beta)\, v_{x+1}^t] e^{i\rho_u} \\ v_x^{t+1} &= [i\sin(\beta)\, u_{x-1}^t + \cos(\beta)\, v_{x-1}^t] e^{i\rho_v}, \end{aligned} \qquad (2)$$

where $u_x^t$ and $v_x^t$ are the complex amplitudes of the two-component wavefunction $|\psi(t)\rangle = \sum_x u_x^t |x\rangle \otimes |\leftarrow\rangle + v_x^t |x\rangle \otimes |\rightarrow\rangle$ at a discrete time step $t$ and lattice position $x$. The two spin-like components $u, v$ propagate left and right, respectively, as illustrated in the mesh lattice in Fig. 2A. The coupling between them is set by the beam splitting parameter $\beta$. A complex lattice potential $\rho_{u,v} = \varphi_{u,v} - ig_{u,v}$ can be applied, also depending on time and position. In this photonic lattice, the real potential $\varphi_{u,v}$ corresponds to an additional phase that is picked up during propagation while the non-Hermitian part $g$ corresponds to optical gain ($g > 0$) and loss ($g < 0$). The system is periodic in both time and space, with a spatiotemporal unit cell extending over two steps in space as well as two steps in time, as indicated by the shaded area in Fig. 2A.

We show that this system may indeed exhibit momentum band winding: Selecting, for instance, $\beta = \pi/4$, $\varphi_{u,v} = 0$ and $g_{u,v} = 0.03$, we can derive both the energy and the momentum band structure. Examining the energy band structure first, as depicted in Fig. 2C-E, we can see that there is no winding of the energy band and hence no signs of non-trivial topology. However, a different situation is uncovered when instead the momentum band structure is considered, as shown in Fig. 2F-H: The momentum band structure looks entirely distinct from the energy band structure, with no immediately evident relation between the two. Moreover, examining the distribution of momentum eigenvalues on the complex plane, we observe that there is indeed a momentum band winding (or more precisely, a winding of the transfer matrix eigenvalues as shown in Fig. 2C) and consequently non-trivial topology, which is evident only from the complex momentum band structure yet hidden from conventional considerations of the energy band structure. The sign of this winding number is controlled by the sign of $g$, meaning in this case two distinct phases with opposite windings exist. Features of the momentum band structure are tightly related to physical phenomenology in time, as seen in prior work (*4, 9, 13*) and also detailed in Section 2 of the Supplementary Text. Thus, systems with opposite momentum band winding may be combined to form a topological interface in time, as depicted in Fig. 3A.

*Results.* Experimentally, photonic quantum walks may be implemented using coupled optical fiber loops (*26–29*). Here, the length difference between two optical fibers coupled by a beam splitter, like in our setup illustrated in Fig. 2B, introduces a synthetic spatial dimension encoded by virtue of temporal multiplexing. Light in this experiment undergoes a quantum walk on a spatiotemporal lattice which may be measured at every site and tailored using optical modulators (see Materials and Methods). By employing an interplay of optical amplification and time-dependent intensity modulation, non-Hermiticity tailorable in both space and time is implemented.

By abruptly switching the optical gain and loss we create a topological interface in time across which the momentum band winding number changes. In Fig. 3B the resulting light propagation is shown: Observing the intensity profile along the time axis, a clear exponential localization in time exactly at the topological time interface can be seen. These experimental results are in line with the predicted localization of temporal momentum eigenstates, i.e., the eigenstates of the momentum operator, due to the momentum band winding in the presence of a topological interface (see Section 5 of the Supplementary Text), akin to the spatial non-Hermitian skin effect. We note that while this effect is independent of the chosen excitation, here we chose to excite the lattice with two Gaussians with opposite group velocities to verify the continued presence of spatial energy transport and that temporal localization occurs independently of spatial position. The non-Hermitian protocol to generate this excitation is detailed in the Materials and Methods.

Perhaps the most striking feature of topological effects in physics is their robustness to disorder. In increasingly disordered systems, where system parameters are more and more randomly perturbed, topological gaps shrink and eventually close, transitioning the system into topological triviality and erasing any topological phenomenology. All currently known topological physics follows this rule, from Hermitian topological band insulators (*30*) over the non-Hermitian skin effect (*22*) to the recently discovered time-topological edge state (*14*). Studying the behavior of momentum band winding topology under disorder, we see it is counter-intuitively able to break this longstanding law. To this end, we apply disorder *via* a random potential through spatiotemporal randomization of the phases, i.e., $\varphi_{u,v} = \delta(t,x)$ with each $\delta(t,x)$ selected at random from the interval $[-D, D]$. The temporal localization can be quantified by calculating the second moment (*31*) $M_2 = \sum_t (t - \bar{t})^2 \langle \psi(t) | \psi(t) \rangle$ of the intensity distribution, where $\bar{t}$ the cutoff time for the propagation. The calculations shown in Fig. 4A show that the second moment remains largely constant even in the face of arbitrarily large increases of disorder. Numerical calculations of the momentum band winding number shown in the same graph confirm that not just the second moment, but the topology itself also remains not just robust but entirely immune to the disorder increasing. The magnitude of the imaginary part of the momentum spectrum is lower bounded by $|g|$ and remains so under disorder, preserving the winding. These random potentials are likely the most common form of disorder. Given their ubiquity and the fact that sufficiently strong disorder would render any other topological system trivial, the immunity observed here is particularly remarkable. Such disorder arises from common issues in a variety of systems such as for example sound perturbing fiber-optical connections (*32*), imprecisely tuned microwave resonators (*33*) or charge-donating substrate impurities in 2D materials (*34*), to name just a few. Even so, we also investigate exotic non-Hermitian disorder, specifically spatiotemporally random non-Hermiticity through randomly perturbed gain/loss $g_{u,v} = g + \gamma(t,x)$ with each $\gamma(t,x)$ selected at random from the interval $[-G, G]$. The resulting second moments and winding numbers with increasing disorder are presented in Fig. 4B. Here, an eventual transition into a topologically trivial regime which coincides with a vanishing second moment, and thus temporal delocalization can be observed. Nevertheless, we note that the transition occurs only at extreme fluctuations corresponding to randomized gain with a peak value of over 40% per time step, where the randomized gain $g + \gamma(t,x)$ can far exceed the disorder-free gain parameter $g$ by up to over 1000%.

In order to verify the disorder-immunity of the topological temporal localization experimentally, we use fast amplitude and phase modulators (see Materials and Methods) to observe propagation under disorder. Firstly, applying moderate spatiotemporally random non-Hermiticity, we observe that the light field, as shown in Fig. 2C, is very similar to the disorder-free case shown, with exponential localization in time clearly visible. Secondly, the propagation with the maximal real disorder potential additionally applied is presented in Fig. 2D. It can directly be seen that the topological temporal localization persists, showing that it is immune to disorder with arbitrarily large values of disorder strength $D$. Note that in space, Anderson localisation due to the disorder has set in. Finally, we show an example of the effect breaking down due to extreme spatiotemporally randomised non-Hermiticity in Fig. 2E. We point out that this result is taken from a simulation, as the extreme amplifications and associated powers required to cause the topology to become trivial were beyond the technical capabilities of available optical amplifiers and the detectors in the experiment. Only under these extreme conditions, the localization breaks down.

*Conclusion.* We have proposed and experimentally realized in time-varying photonic quantum walks the winding topology of complex momentum bands hidden from the energy band structure and, as a direct consequence, topological temporal localization at a time interface. We further observed and verified in experiment that, remarkably and unlike any other topological effect, this type of topology and temporal localization are immune to the ubiquitous real disorder and only break down under extreme spatiotemporally random non-Hermiticity. These intriguing results invite the deeper exploration of momentum band physics in general and also more specifically the complex momentum band structure: As particularly exciting we highlight the perspective of spatiotemporally crystalline systems (*7*, *35*, *36*), where due to a combined energy-momentum topology rich phenomenology in a dimensional interplay of space and time may emerge. Additionally, the inspiring recent progress on intricate topological structures beyond winding such as knots (*37*) gives rise to the compelling question of their existence and possible consequences in momentum or even energy-momentum band structure. Moreover, as the physics described in this paper is not at all specific to our platform, we expect our results to both catalyze and be catalyzed by the rapid recent efforts in engineering time interfaces and time-varying systems, for instance in optically nonlinear media (*38*, *39*), ultracold atoms (*40*), microwave circuits (*41*, *42*), fluids (*43*) or acoustics (*44*, *45*). Hence, the general nature and desirable properties of this type of topological physics together has the potential to be greatly useful to various broadly useful technologies by enabling, for example, ultrarobust topological lasing (*46*, *47*), pulse shaping (*48*, *49*) or amplification (*50*).

**References and Notes**

1. N. W. Ashcroft, N. D. Mermin, *Solid State Physics* (Holt, Rinehart and Winston, New York, 1976).

2. F. Wilczek, Crystals in time. *Scientific American* **321**, 28–36 (2019).

3. M. P. Zaletel, M. Lukin, C. Monroe, C. Nayak, F. Wilczek, N. Y. Yao, *Colloquium* : Quantum and classical discrete time crystals. *Rev. Mod. Phys.* **95**, 031001 (2023).

4. E. Galiffi, R. Tirole, S. Yin, H. Li, S. Vezzoli, P. A. Huidobro, M. G. Silveirinha, R. Sapienza, A. Alù, J. Pendry, Photonics of time-varying media. *Advanced Photonics* **4**, 014002–014002 (2022).

5. N. Engheta, Four-dimensional optics using time-varying metamaterials. *Science* **379**, 1190–1191 (2023).

6. M. F. Saleh, A. Armaroli, T. X. Tran, A. Marini, F. Belli, A. Abdolvand, F. Biancalana, Raman-induced temporal condensed matter physics in gas-filled photonic crystal fibers. *Opt. Express* **23**, 11879 (2015).

7. J. Feis, S. Weidemann, T. Sheppard, H. M. Price, A. Szameit, Space-time-topological events in photonic quantum walks. *Nat. Photon.* **19**, 518–525 (2025).

8. M. M. Asgari, P. Garg, X. Wang, M. S. Mirmoosa, C. Rockstuhl, V. Asadchy, Theory and applications of photonic time crystals: a tutorial. *Adv. Opt. Photon.* **16**, 958 (2024).

9. E. Lustig, O. Segal, S. Saha, C. Fruhling, V. M. Shalaev, A. Boltasseva, M. Segev, Photonic time-crystals - fundamental concepts [Invited]. *Opt. Express* **31**, 9165 (2023).

10. J. T. Mendonça, P. K. Shukla, Time Refraction and Time Reflection: Two Basic Concepts. *Phys. Scr.* **65**, 160–163 (2002).

11. J. Reyes-Ayona, P. Halevi, Observation of genuine wave vector (k or β) gap in a dynamic transmission line and temporal photonic crystals. *Applied Physics Letters* **107** (2015).

12. A. M. Shaltout, J. Fang, A. V. Kildishev, V. M. Shalaev, "Photonic Time-Crystals and Momentum Band-Gaps" in *Conference on Lasers and Electro-Optics* (OSA, San Jose, California, 2016), p. FM1D.4.

13. E. Lustig, Y. Sharabi, M. Segev, Topological aspects of photonic time crystals. *Optica* **5**, 1390 (2018).

14. Y. Ren, K. Ye, Q. Chen, F. Chen, L. Zhang, Y. Pan, W. Li, X. Li, L. Zhang, H. Chen, Y. Yang, Observation of momentum-gap topology of light at temporal interfaces in a time-synthetic lattice. *Nat. Commun.* **16**, 707 (2025).

15. Y. Pan, M.-I. Cohen, M. Segev, Superluminal k -Gap Solitons in Nonlinear Photonic Time Crystals. *Phys. Rev. Lett.* **130**, 233801 (2023).

16. M. Lyubarov, Y. Lumer, A. Dikopoltsev, E. Lustig, Y. Sharabi, M. Segev, Amplified emission and lasing in photonic time crystals. *Science* **377**, 425–428 (2022).

17. M. Z. Hasan, C. L. Kane, Colloquium : Topological insulators. *Rev. Mod. Phys.* **82**, 3045–3067 (2010).

18. M. S. Rudner, N. H. Lindner, Band structure engineering and non-equilibrium dynamics in Floquet topological insulators. *Nat. Rev. Phys.* **2**, 229–244 (2020).

19. Z. Wang, Y. Chong, J. D. Joannopoulos, M. Soljačić, Observation of unidirectional backscattering-immune topological electromagnetic states. *Nature* **461**, 772–775 (2009).

20. M. C. Rechtsman, J. M. Zeuner, Y. Plotnik, Y. Lumer, D. Podolsky, F. Dreisow, S. Nolte, M. Segev, A. Szameit, Photonic Floquet topological insulators. *Nature* **496**, 196–200 (2013).

21. S. Yao, Z. Wang, Edge States and Topological Invariants of Non-Hermitian Systems. *Phys. Rev. Lett.* **121**, 086803 (2018).

22. S. Weidemann, M. Kremer, T. Helbig, T. Hofmann, A. Stegmaier, M. Greiter, R. Thomale, A. Szameit, Topological funneling of light. *Science* **368**, 311–314 (2020).

23. J. Xiong, X. Zhang, L. Duan, J. Wang, Y. Long, H. Hou, L. Yu, L. Zou, B. Zhang, Observation of wave amplification and temporal topological state in a non-synthetic photonic time crystal. *Nat. Commun.* **16**, 11182 (2025).

24. J. J. Sakurai, J. Napolitano, *Modern Quantum Mechanics* (Cambridge University Press, Cambridge, Second edition., 2017).

25. A. F. Beardon, *Complex Analysis: The Argument Principle in Analysis and Topology* (Wiley, Chichester [Eng.]; New York, 1979) *A Wiley-Interscience publication*.

26. A. Schreiber, K. N. Cassemiro, V. Potoček, A. Gábris, P. J. Mosley, E. Andersson, I. Jex, Ch. Silberhorn, Photons Walking the Line: A Quantum Walk with Adjustable Coin Operations. *Phys. Rev. Lett.* **104**, 050502 (2010).

27. A. Regensburger, C. Bersch, M.-A. Miri, G. Onishchukov, D. N. Christodoulides, U. Peschel, Parity–time synthetic photonic lattices. *Nature* **488**, 167–171 (2012).

28. A. L. Marques Muniz, F. O. Wu, P. S. Jung, M. Khajavikhan, D. N. Christodoulides, U. Peschel, Observation of photon-photon thermodynamic processes under negative optical temperature conditions. *Science* **379**, 1019–1023 (2023).

29. A. F. Adiyatullin, L. K. Upreti, C. Lechevalier, C. Evain, F. Copie, P. Suret, S. Randoux, P. Delplace, A. Amo, Topological Properties of Floquet Winding Bands in a Photonic Lattice. *Phys. Rev. Lett.* **130**, 056901 (2023).

30. J. K. Asbóth, L. Oroszlány, A. Pályi, *A Short Course on Topological Insulators* (Springer International Publishing, Cham, 2016) vol. 919 of *Lecture Notes in Physics*.

31. B. Kramer, A. MacKinnon, Localization: theory and experiment. *Rep. Prog. Phys.* **56**, 1469–1564 (1993).

32. L.-S. Ma, P. Jungner, J. Ye, J. L. Hall, Delivering the same optical frequency at two places: accurate cancellation of phase noise introduced by an optical fiber or other time-varying path. *Opt. Lett.* **19**, 1777 (1994).

33. J. Van Damme, S. Massar, R. Acharya, Ts. Ivanov, D. Perez Lozano, Y. Canvel, M. Demarets, D. Vangoidsenhoven, Y. Hermans, J. G. Lai, A. M. Vadiraj, M. Mongillo, D. Wan, J. De Boeck, A. Potočnik, K. De Greve, Advanced CMOS manufacturing of superconducting qubits on 300 mm wafers. *Nature* **634**, 74–79 (2024).

34. Y. Zhang, V. W. Brar, C. Girit, A. Zettl, M. F. Crommie, Origin of spatial charge inhomogeneity in graphene. *Nature Phys* **5**, 722–726 (2009).

35. Y. Sharabi, A. Dikopoltsev, E. Lustig, Y. Lumer, M. Segev, Spatiotemporal photonic crystals. *Optica* **9**, 585–592 (2022).

36. O. Segal, Y. Plotnik, E. Lustig, Y. Sharabi, M.-I. Cohen, A. Dikopoltsev, M. Segev, Two-Dimensional Topological Edge States in Periodic Space-Time Interfaces. *Phys. Rev. Lett.* **135**, 163801 (2025).

37. K. Wang, A. Dutt, C. C. Wojcik, S. Fan, Topological complex-energy braiding of non-Hermitian bands. *Nature* **598**, 59–64 (2021).

38. R. Tirole, S. Vezzoli, E. Galiffi, I. Robertson, D. Maurice, B. Tilmann, S. A. Maier, J. B. Pendry, R. Sapienza, Double-slit time diffraction at optical frequencies. *Nat. Phys.* **19**, 999–1002 (2023).

39. K. Pang, M. Z. Alam, Y. Zhou, C. Liu, O. Reshef, K. Manukyan, M. Voegtle, A. Pennathur, C. Tseng, X. Su, H. Song, Z. Zhao, R. Zhang, H. Song, N. Hu, A. Almaiman, J. M. Dawlaty, R. W. Boyd, M. Tur, A. E. Willner, Adiabatic Frequency Conversion Using a Time-Varying Epsilon-Near-Zero Metasurface. *Nano Lett.* **21**, 5907–5913 (2021).

40. Z. Dong, H. Li, T. Wan, Q. Liang, Z. Yang, B. Yan, Quantum time reflection and refraction of ultracold atoms. *Nat. Photon.* **18**, 68–73 (2024).

41. X. Wang, M. S. Mirmoosa, V. S. Asadchy, C. Rockstuhl, S. Fan, S. A. Tretyakov, Metasurface-based realization of photonic time crystals. *Science Advances* **9**, eadg7541 (2023).

42. T. R. Jones, A. V. Kildishev, M. Segev, D. Peroulis, Time-reflection of microwaves by a fast optically-controlled time-boundary. *Nat. Commun.* **15**, 6786 (2024).

43. V. Bacot, M. Labousse, A. Eddi, M. Fink, E. Fort, Time reversal and holography with spacetime transformations. *Nature Physics* **12**, 972–977 (2016).

44. S. Tong, Q. Zhang, G. Li, K. Zhang, C. Qiu, Acoustic realization of monoatomic topological space-time crystals. *Newton*, 100304 (2025).

45. Z. Liu, X. Zhu, Z. G. Zhang, W. M. Zhang, X. Chen, Y. Q. Yang, R. W. Peng, M. Wang, J. Li, H. W. Wu, Direct Observation of k-Gaps in Dynamically Modulated Phononic Time Crystal. arXiv [Preprint] (2025). https://doi.org/10.48550/ARXIV.2505.07160.

46. M. A. Bandres, S. Wittek, G. Harari, M. Parto, J. Ren, M. Segev, D. N. Christodoulides, M. Khajavikhan, Topological insulator laser: Experiments. *Science* **359**, eaar4005 (2018).

47. B. Bahari, A. Ndao, F. Vallini, A. El Amili, Y. Fainman, B. Kanté, Nonreciprocal lasing in topological cavities of arbitrary geometries. *Science* **358**, 636–640 (2017).

48. A. G. Löhr, M. Y. Ivanov, M. A. Khokhlova, Controlled compression, amplification and frequency up-conversion of optical pulses by media with time-dependent refractive index. *Nanophotonics* **12**, 2921–2928 (2023).

49. E. Galiffi, G. Xu, S. Yin, H. Moussa, Y. Ra'di, A. Alù, Broadband coherent wave control through photonic collisions at time interfaces. *Nat. Phys.* **19**, 1703–1708 (2023).

50. E. Galiffi, P. A. Huidobro, J. B. Pendry, Broadband Nonreciprocal Amplification in Luminal Metamaterials. *Phys. Rev. Lett.* **123**, 206101 (2019).

51. M.-A. Miri, A. Regensburger, U. Peschel, D. N. Christodoulides, Optical mesh lattices with PT symmetry. *Phys. Rev. A* **86**, 023807 (2012).

52. T. Eichelkraut, R. Heilmann, S. Weimann, S. Stützer, F. Dreisow, D. N. Christodoulides, S. Nolte, A. Szameit, Mobility transition from ballistic to diffusive transport in non-Hermitian lattices. *Nat Commun* **4**, 2533 (2013).

53. M. Wimmer, A. Regensburger, M.-A. Miri, C. Bersch, D. N. Christodoulides, U. Peschel, Observation of optical solitons in PT-symmetric lattices. *Nat Commun* **6**, 7782 (2015).

54. Y. Tanaka, A constructive approach to topological invariants for one-dimensional strictly local operators. *Journal of Mathematical Analysis and Applications* **500**, 125072 (2021).

55. P. Yeh, *Optical Waves in Layered Media* (Wiley-Interscience, Hoboken, NJ, 2005)*Wiley series in pure and applied optics*.

56. Z. Gong, Y. Ashida, K. Kawabata, K. Takasan, S. Higashikawa, M. Ueda, Topological Phases of Non-Hermitian Systems. *Phys. Rev. X* **8**, 031079 (2018).

57. S. Weidemann, M. Kremer, S. Longhi, A. Szameit, Topological triple phase transition in non-Hermitian Floquet quasicrystals. *Nature* **601**, 354–359 (2022).

58. R. L. Burden, J. D. Faires, A. M. Burden, *Numerical Analysis* (Cengage Learning, Boston, MA, Tenth edition., 2016).

**Acknowledgments:** We thank Dominik Heß for assistance with figure design.

**Funding:**

    Royal Society grant UF160112 (TS, HMP)

    Royal Society grant URF\R\221004 (TS, HMP)

    Royal Society grant RGF\EA\180121 (TS, HMP)

    Royal Society grant RGF\R1\180071 (TS, HMP)

    Engineering and Physical Sciences Research Council grant EP/W016141/1 (TS, HMP)

    Engineering and Physical Sciences Research Council grant EP/Y01510X/1 (TS, HMP)

    Engineering and Physical Sciences Research Council grant UKRI2226 (TS, HMP)

    Deutsche Forschungsgemeinschaft grant SZ 276/9-2 (AS)

    Deutsche Forschungsgemeinschaft grant SZ 276/19-1 (AS)

    Deutsche Forschungsgemeinschaft grant SZ 276/21-1 (AS)

    Deutsche Forschungsgemeinschaft grant GRK 2676/1-2023 'Imaging of Quantum Systems' (project no. 437567992) (AS)

    Deutsche Forschungsgemeinschaft SFB 1477 'Light-Matter Interactions at Interfaces' (project no. 441234705) (AS)

    Krupp von Bohlen and Halbach Foundation (AS)

    Leverhulme Trust Study Abroad Studentship (JF)

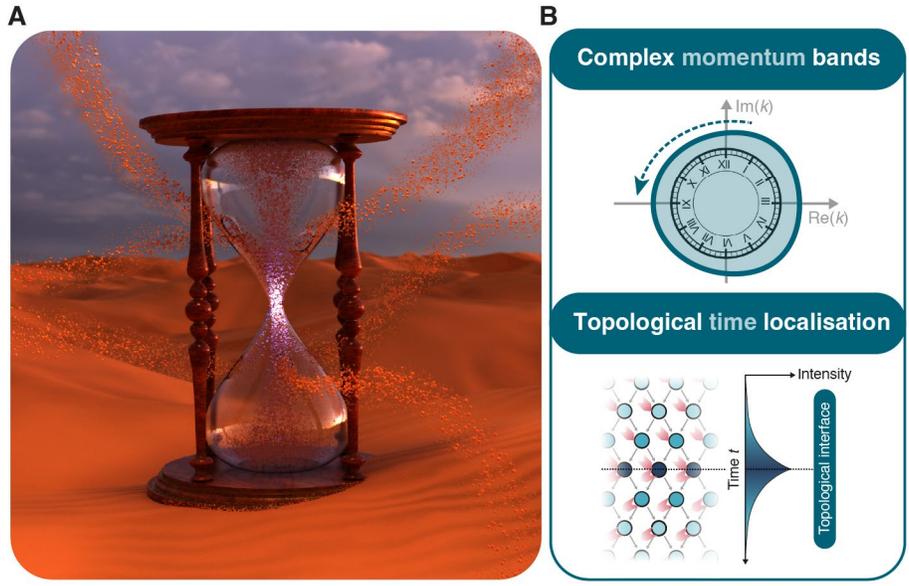

**Fig. 1. Momentum band winding topology.** **(A)** Conceptual illustration of topological temporal localization. **(B)** Complex momentum bands may exhibit winding topology, which can lead to topological localization in time at a time interface, where a temporally crystalline structure is abruptly switched.

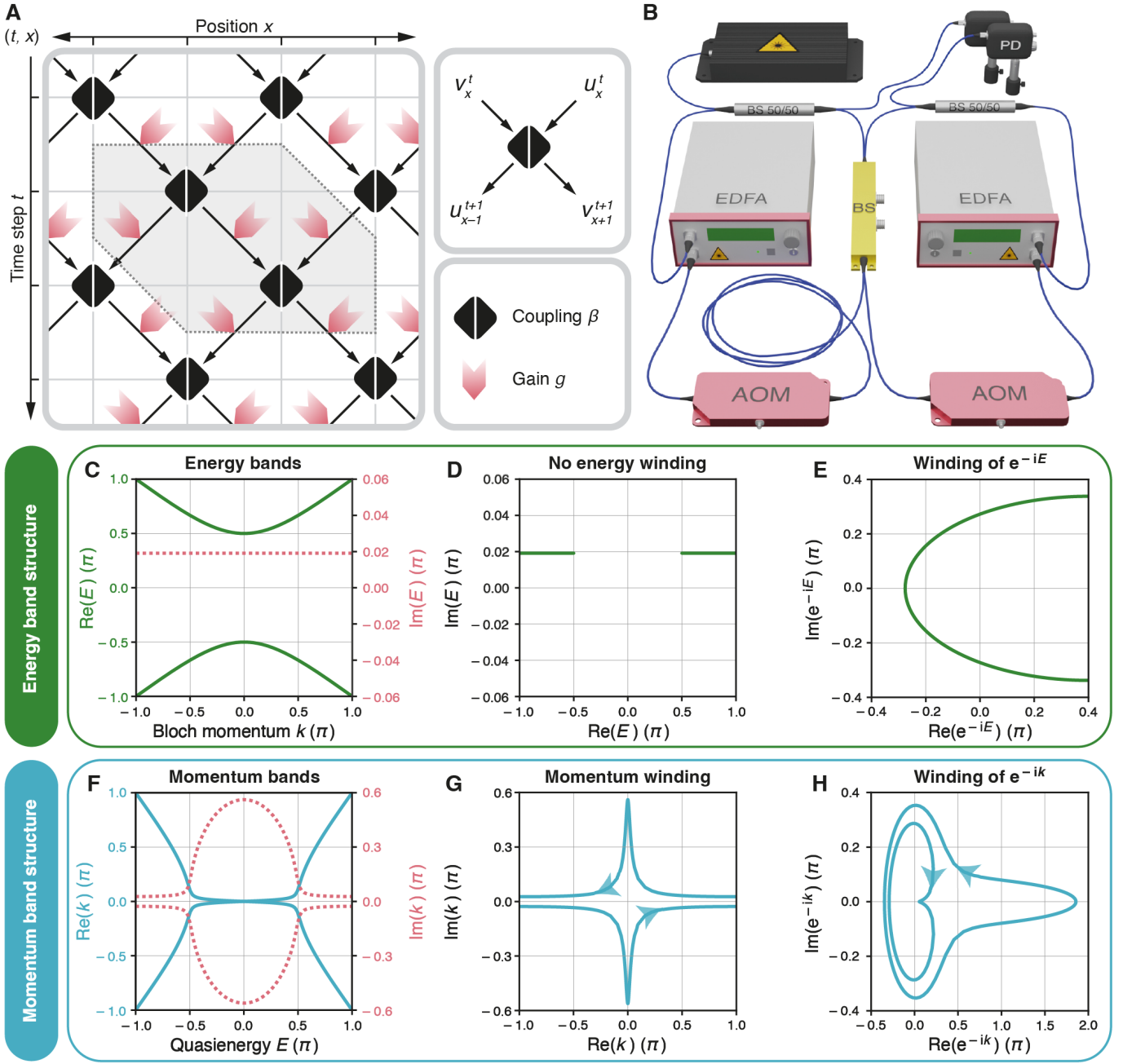

**Fig. 2. Photonic quantum walks with winding momentum bands.** (**A**) The photonic quantum walk takes place on a space-time-periodic lattice with beam splitters coupling the two spin-like components of the wavefunction and spatiotemporally adjustable non-Hermiticity through a gain $g = 0.03$. The spatiotemporal unit cell is shaded in gray. (**B**) Simplified illustration of the experiment realizing photonic quantum walks. Two loops of optical fiber are connected *via* a beam splitter (BS). The non-Hermitian modulation is enabled by acousto-optical amplitude modulators (AOMs) and erbium-doped fiber amplifiers (EDFAs). (**C**)-(**E**) Complex energy bands of the quantum walk shown as a function of momentum in (C), in the complex plane in (D) and as eigenvalues $\exp(-iE)$ of the time evolution operator in (E). No winding is visible. (**F**)-(**G**), Complex momentum bands of the quantum walk shown as a function of energy in (F), in the complex plane in (G) and as eigenvalues $\exp(-ik)$ of the transfer matrix in (H), where a winding can be seen.

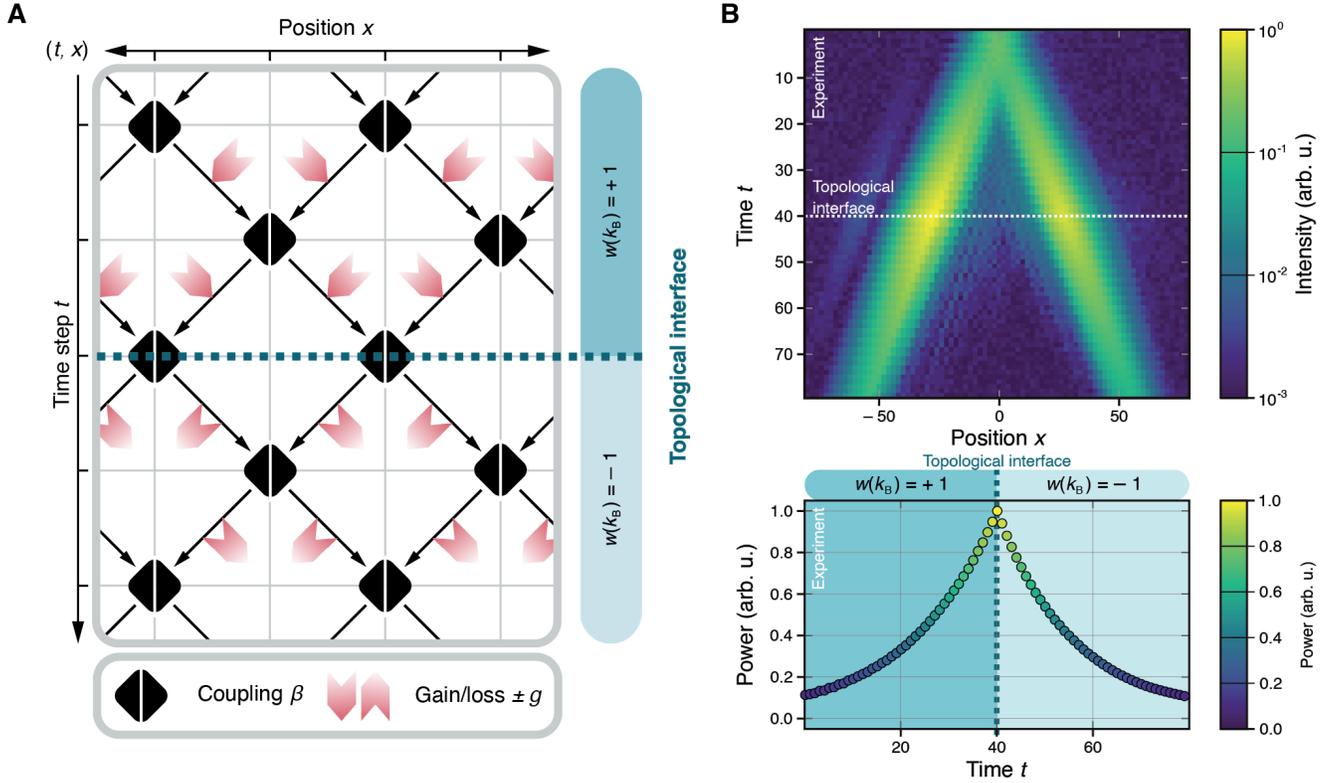

**Fig. 3. Observation of topological localisation in time from momentum band winding. (A)** Through an abrupt temporal switching of gain and loss, a time interface is created across which the momentum band winding number and thus the topology changes. The winding has been calculated with respect to a base point $k_B = 0$. **(B)** Localisation is mandated by the change in topology at the time interface. We excite the system with two tilted Gaussians and see (top) that such localisation in time indeed occurs, independent of spatial location while spatial energy transport persists. Examining the profile of the total power (bottom) shows the exponential nature of the topological temporal localisation. The experiment has been performed with a gain/loss strength of $g = \pm 0.03$.

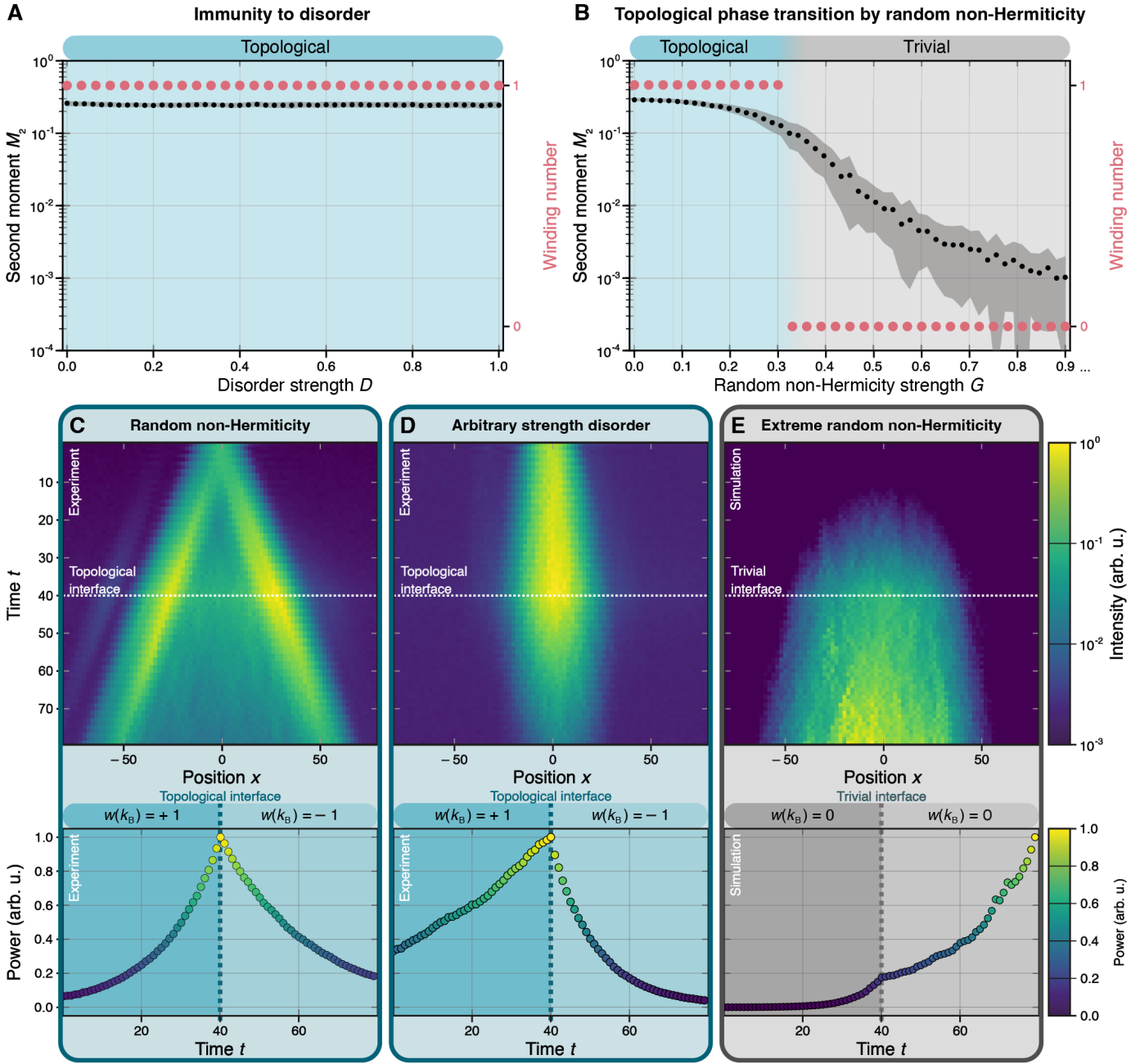

**Fig. 4. Momentum band winding topology under disorder.** (**A**) Topology and localisation remain immune against arbitrarily strong disorder applied through random potentials as evidenced by constant momentum band winding number and second moment, respectively. (**B**) Only extreme spatiotemporally random non-Hermiticity is able to induce a topological phase transition into a topologically trivial regime. This coincides with a delocalisation transition, evidenced by a drop-off of the second moment. (**C**) For moderate spatiotemporally random non-Hermiticity ($D = 0, G = 0.18$), the propagation and topological localisation is highly robust and remains essentially qualitatively unchanged to the disorder-free case. (**D**) Under arbitrarily strong disorder through random potentials ($D = \pi, G = 0.18$), the topological localisation clearly persists, verifying the immunity of the topology. In space, Anderson localisation has set in. (**E**) Only for extreme spatiotemporally random non-Hermiticity ($G = 0.4$) the topological localisation breaks down.

15